\documentclass[prl,aps,twocolumn,groupedaddress]{revtex4}
\begin{document}

\title{\bf Supersymmetry and Goldstino-like Mode in Bose-Fermi Mixtures}

\author{Yue Yu$^1$}

\author{Kun Yang$^{2}$}

\affiliation{$^1$Institute of Theoretical Physics, Chinese Academy of
Sciences, P.O. Box 2735, Beijing 100080, China}

\affiliation{$^2$NHMFL and Department of Physics, Florida State
University, Tallahassee, Florida 32306, USA}

\date{\today}
\begin{abstract}
Supersymmetry is assumed to be a basic symmetry of the world in many
high energy theories, but none of the super partners of any known
elementary particle has been observed yet. We argue that
supersymmetry can also be realized and studied in ultracold atomic
systems with a mixture of bosons and fermions, with properly tuned
interactions and single particle dispersion. We further show that in
such non-releativistic systems supersymmetry is either spontaneously
broken, or explicitly broken by a chemical potential difference
between the bosons and fermions. In both cases the system supports a
{\em sharp} fermionic collective mode similar to the
\emph{Goldstino} mode in high-energy physics, due to supersymmetry. We also discuss possible
ways to detect this mode experimentally.
\end{abstract}


\maketitle

Supersymmetry, which is a symmetry that relates bosons and
fermions, became a topic of strong interest in elementary particle
physics after Wess and Zumino constructed the first ``realistic"
model \cite{wz}. It may play a fundamental role because the
supersymmetry algebra is the only graded Lie algebra of the
symmetries of the $S$-matrix consistent with relativistic quantum
field theory \cite{hsl}. The supersymmetric string theory is a
unique theory expected to give a unified description of all
interactions in nature \cite{gsw}. However, none of the super
partners (which have identical properties except for opposite
statistics) of any known elementary particles has been found in
experiments thus far. Therefore, it is extremely important to
study the breaking of the supersymmetry.

Recent experimental progress in mixtures of ultracold Bose and
Fermi atoms \cite{exp} provides an opportunity to realize and
study supersymmetry in such atomic systems. Theoretically, several
different models that exhibit supersymmetry have been studied. An ultracold superstring model was constructed
\cite{th1}. The physical behavior of an exactly soluable model of
one-dimensional Bose-Fermi mixture was investigated \cite{th2}. A
general formalism to study such supersymmetric systems based on
coherent state path integral was set up in Ref.
\onlinecite{mieck}. In a recent work \cite{th3}, we studied a
supersymmetric Hubbard model and focused on the Mott insulator
phase for bosons.

In the present paper we study some general properties of
supersymmetric Bose-Fermi mixtures, in which bosons and fermions are
supersymmetric partners of each other. We show that in the presence
of time-reversal symmetry, supersymmetry is always broken, either
spontaneously or by a chemical potential difference between bosons
and fermions. For a truly supersymmetric grand canonical
Hamiltonian, we find the system contains bosons only even though the
grand canonical Hamiltonian is invariant under a supersymmetry
transformation that turns a boson into a fermion; in this case
supersymmetry is spontaneously broken. To support a finite density
of fermions, one needs a
chemical potential difference between the bosons and fermions:
$\Delta \mu = \mu_F-\mu_B > 0$, which breaks supersymmetry of grand
canonical Hamiltonian explicitly but keeps the canonical Hamiltonian
supersymmetric. We find in both cases the system
supports a sharp fermionic collective excitation similar to the Goldstino mode
in supersymmetric high energy theories,
which is gapless in the former case while has a gap equal to $\Delta
\mu$ for the latter. For simplicity we refer to this Goldstino-like mode as ``Goldstino"
in the following. We will also discuss its possible experimental
detection.

While our results are general, in the following we illustrate them
by considering a simple lattice model (in the grand canonical
ensemble) with mixture between a single species of bosons and a single
species of fermions:
\begin{eqnarray}
\label{HG}
&&H_G=H-\mu_F N_F-\mu_B N_B;\\
&&H=\hat{T}+\hat{V};\\
&&\hat{T}=-\sum_{i\ne j}(t^B_{ij}a^\dagger_ia_j+t^F_{ij}f^\dag_i
f_j); \\
&&\hat{V}=\sum_{(ij)}[U^{BB}_{ij}
n^a_in^a_j+U^{BF}_{ij}n^a_in^f_j+U^{FF}_{ij}n^f_in^f_j].
\end{eqnarray}
Here $a_i$ and $f_i$ are the boson and fermion operators on site
$i$, and $n^a_i$ and $n^f_i$ are the corresponding number operators;
$N_B=\sum_i n^a_i$ and $N_F=\sum_i n_i^f$. In the presence of
time-reversal symmetry, all the hopping matrix elements ($t$'s) are
real; we further assume they are all non-negative so the hoppings
are not frustrated as is usually the case.

We now introduce generators of supersymmetry\cite{kuklov}:
\begin{equation}
\label{Q} Q=\sum_i a_i^\dag f_i; \hskip 0,5 cm Q^\dag=\sum_i a_i
f_i^\dag.
\end{equation}
It is easy to verify that they satisfy the following relations:
$Q^2=(Q^\dagger)^2=0$; $\{Q, Q^\dagger\}=N=N_B+N_F$; and $[Q,
N]=[Q^\dagger, N]=0$. From these relation it is clear that $Q$ is a
{\em fermionic} operator. Physically $Q$ turns a fermion into a
boson, and $Q^\dagger$ does the opposite. While in (\ref{Q}) $Q$ is
defined in 2nd quantized notation, we also need to know its
operation on a 1st quantized wave function
\begin{equation}
\psi({\bf x}_1, \cdots, {\bf x}_{N_B}; {\bf y}_1, \cdots, {\bf
y}_{N_F})
\end{equation}
where ${\bf x}$ and ${\bf y}$ are the coordinates of the bosons and
fermions respectively, so that $\psi$ is symmetric under the
exchange of ${\bf x}$'s while antisymmetric under the exchange of
${\bf y}$'s. We find (up to a normalization constant)
\begin{eqnarray}
Q\psi({\bf x}_1, \cdots, {\bf y}_{N_F})&=&S_{{\bf x}_1, \cdots,
{\bf x}_{N_B}, {\bf y}_1}\psi({\bf x}_1, \cdots, {\bf y}_{N_F}),\\
Q^\dag\psi({\bf x}_1, \cdots, {\bf y}_{N_F})&=&A_{{\bf x}_1, {\bf
y}_1,\cdots, {\bf y}_{N_F}}\psi({\bf x}_1, \cdots, {\bf y}_{N_F}),
\end{eqnarray}
where $S_{{\bf x}_1, \cdots, {\bf x}_{N_B}, {\bf y}_1}$ is the
symmetrization operator for coordinates ${\bf x}_1, \cdots, {\bf
x}_{N_B}, {\bf y}_1$, and $A_{{\bf x}_1, {\bf y}_1,\cdots, {\bf
y}_{N_F}}$ is the antisymmetrization operator for coordinates ${\bf
x}_1, {\bf y}_1,\cdots, {\bf y}_{N_F}$. Physically, $S$ turns a
fermion into a boson by symmetrizing one fermion coordinate with
respect to all boson coordinates, and $A$ turns a boson into a
fermion by antisymmetrizing one boson coordinate with respect to all
fermion coordinates.

If the bosons and fermions have the same dispersion and interaction
strength, namely $t^B_{ij}=t^F_{ij}$, and
$U^{BB}_{ij}=U^{BF}_{ij}=U^{FF}_{ij}$, it is easy to show that the
canonical ensemble Hamiltonian $H$ is supersymmetric, i.e.,
\begin{equation}
[Q,H]=[Q^\dagger,H]=0.
\end{equation}
For completeness we also present a generic example of
supersymmetric $H$ in the continuum, in 1st quantization:
\begin{eqnarray}
H'_{1st}&=&\sum_{i\le N_B}[-\nabla_{{\bf x}_i}^2+V({\bf x}_i)]+\sum_{i\le
N_F}[-\nabla_{{\bf y}_i}^2+V({\bf y}_i)]\nonumber\\
&+&\sum_{i<j\leq N_B}U({\bf x}_i-{\bf x}_j)+\sum_{i<j\leq N_F}U({\bf
y}_i-{\bf y}_j)\nonumber\\
&+&\sum_{i\leq N_B,j\leq N_F}U({\bf x}_i-{\bf y}_j),
\end{eqnarray}
where $V$ is single particle potential and $U$ is two-body
interaction. Of course the 2nd quantized, lattice Hamiltonian (2) also has a
corresponding 1st quantized form:
\begin{equation}
H_{1st}=\hat{T}_{1st}+\hat{V}_{1st},
\end{equation}
where the form of $\hat{T}_{1st}$ and $\hat{V}_{1st}$ can be deduced from Eqs. (3) and (4).
If $\psi({\bf x}_1, \cdots, {\bf y}_{N_F})$ is an eigen
wave function of the Hamiltonian $H_{1st}$ with $N_B$ bosons and $N_F$
fermions, then $Q\psi({\bf x}_1, \cdots, {\bf y}_{N_F})$ is an eigen
wave functionof $H_{1st}$ with $N_B+1$ bosons and $N_F-1$ fermions with
exactly the same eigen energy. Similarly $Q^\dag\psi({\bf x}_1,
\cdots, {\bf y}_{N_F})$ is an eigen wave function of $H_{1st}$ with $N_B-1$
bosons and $N_F+1$ fermions, also with exactly the same eigen
energy.

If we further have the same chemical potential for the bosons and
fermions: $\mu_F=\mu_B$, the grand canonical Hamiltonian is also
supersymmetric:
\begin{equation}
[Q,H_G]=[Q^\dagger,H_G]=0.
\label{susy}
\end{equation}
Given the great tunability of parameters in cold atom systems, we
expect such conditions can be reached in a variety of systems. In
the following we discuss consequences of such supersymmetry when
present.

We start by considering the case where $H_G$ is supersymmetric:
$[Q,H_G]=0$. In this case we can prove that the ground state of
$H_G$ contains no or only one fermion. Let us {\em assume} the ground
state contains more than one fermion, and has the wave function
$\psi({\bf x}_1, \cdots, {\bf x}_{N_B}; {\bf y}_1, \cdots, {\bf
y}_{N_F})$. Due to the time reversal symmetry of $H_G$, $\psi$ can
be chosen to be real. Now construct a trial state:
\begin{equation}
\tilde{\psi}({\bf x}_1, \cdots, {\bf x}_{N_B}; {\bf y}_1, \cdots,
{\bf y}_{N_F})=|\psi({\bf x}_1, \cdots, {\bf y}_{N_F})|.
\end{equation}
Obviously, $\tilde{\psi}$ is non-negative and different from $\psi$
for $N_F > 1$, as the latter changes sign under the exchange of
${\bf y}$'s. In fact $\tilde{\psi}$ is a two-component or
``spin"-1/2 boson wave function as it is also symmetric under the
exchange of ${\bf y}$'s. Since $\tilde{\psi}$ and $\psi$ differ in phase only, we find the potential/interaction energy does not change:
$\langle\psi|\hat{V}_{1st}|\psi\rangle=\langle\tilde{\psi}|\hat{V}_{1st}|\tilde{\psi}\rangle$, because $\hat{V}_{1st}$ depends only on density but not the phase of wave function. The situation is different for $\hat{T}_{1st}$, which {\em is} sensitive to the wave function phase.
Because $\tilde{\psi}$ and $\psi$ are the same for certain configurations (when $\psi \ge 0$) and differ by a $-$ sign
for others (when $\psi < 0$), we find the expectation value of every term in $\langle \hat{T}_{1st}\rangle$ is either the same for $|\psi\rangle$
and $|\tilde{\psi}\rangle$, or differ by a $-$ sign. Because $\tilde{\psi}$ is non-negative, and all $t$'s in $\hat{T}$ are non-negative (meaning $\hat{T}$ only has negative matrix elements), we find {\em every single term in $\langle\tilde{\psi}|\hat{T}_{1st}|\tilde{\psi}\rangle$ is non-positive}.
We thus have $\langle\psi|\hat{T}_{1st}|\psi\rangle\ge\langle\tilde{\psi}|\hat{T}_{1st}|\tilde{\psi}\rangle$, and as a result
$\langle\psi|H_{1st}|\psi\rangle\ge\langle\tilde{\psi}|H_{1st}|\tilde{\psi}\rangle$.
Since in general $|\tilde{\psi}\rangle$ is {\em not} an eigen state
of $H_{1st}$, and the ground state of such spin-1/2 boson system is spin
fully polarized\cite{lieb,yang}, from variational theorem we find
$\langle\psi|H_G|\psi\rangle
> E_0$, where $E_0$ is the lowest $H_G$ eigen value for the case with $N$ bosons and no fermion
(here we also used the fact that bosons and fermions have the same chemical potential).
This is in contradiction with the assumption that $|\psi\rangle$ is
the ground state of $H_G$. We thus conclude that in the ground state
$N_F$ can only be $0$ or $1$\cite{note}.

We now show that the
ground states have a double degeneracy. Assume $|\psi_0\rangle$ is
the ground state with $N_F=0$. We can then construct a different
state with $N_F=1$ and one fewer boson
\begin{equation}
|\psi_G\rangle=Q^\dagger|\psi_0\rangle,
\end{equation}
which is also an exact eigen state of $H_G$ with exactly the same
energy $E_0$, as guaranteed by supersymmetry (\ref{susy}).
$|\psi_G\rangle$ can be viewed as a zero momentum, fermionic zero
mode of the ground state, which is known as the Goldstino mode in
the high-energy literature.

As a simple example illustrating the results presented above, consider the special case of non-interacting particles with all $U=0$.
For non-interacting bosons we always have $\mu_B=0$ (we measure energy from the bottom of
single particle dispersion) at zero temperature,
regardless of boson number.
Supersymmetry of $H_G$ requires $\mu_F=\mu_B=0$, as a result we can have either no fermion, or
a single fermion occupying the ${\bf k}=0$ state (with zero energy) in the ground state of $H_G$.

It should be clear from the discussion above that not only the
ground states, but {\em all} eigen states of $H_G$ (except for
vacuum with $N=0$) come in degenerate pairs that have the same total
number of particles but differ in fermion number by one due to
(\ref{susy}). The vacuum is special in that it is the {\em only}
state that is supersymmetric, as it is annihilated by {\em both} $Q$
and $Q^\dag$.

We can also construct Goldstino at finite momentum:
\begin{equation}
\label{modes} |\psi_G({\bf q})\rangle=R_{\bf
q}^\dagger|\psi_0\rangle,
\end{equation}
by boosting $Q^\dagger$ to finite momentum:
\begin{equation}
R^\dag_{\bf q}=\sum_i e^{-i{\bf q\cdot x}_i}
a_if^\dag_i=\sum_k a_{\bf k}f^\dag_{\bf k+q}.
\end{equation}
Since the statistics of a single fermion created by $R^\dag_{\bf
q}$ has no physical consequence, $|\psi_G({\bf q})\rangle$ is just
like ferromagnetic spin wave states for $SU(2)$ bosons\cite{yang}; they have exactly the same dispersion and differ only in statistics.
While $|\psi_G({\bf q})\rangle$ is not an exact eigen state of
$H_G$ for finite ${\bf q}$, it approaches one in the long-wave
length limit ${\bf q}\rightarrow 0$, and has quadratic dispersion:
$E_{\bf q}-E_0\propto |{\bf q}|^2$\cite{yang}.

It is appropriate at this point to discuss the relation between the
sharp fermionic collective mode we call Goldstino here, and the Goldstino in high-energy context.
In high-energy context, Goldstino refers to the
Goldstone fermion arising from spontaneous breaking of the global supersymmetry; it is a Weyl spinor with spin-1/2. In our non-relativistic model, the gapless Goldstone fermion mode is also the result of the spontaneous breaking of the global supersymmetry. In this sense, they are quite similar. In fact, if we had a two-component boson system instead of a Bose-Fermi mixture, we would have a pseudospin ferromagnet that breaks SU(2) symmetry, with a branch of gapless spin-wave mode; the spin-wave mode is the Goldstone boson associated with breaking of SU(2) symmetry. In the Bose-Fermi mixture, the second component is fermionic, and the SU(2) symmetry becomes supersymmetry that is generated by the fermionic operator $Q$. The difference here is that the Goldstino is a spinless fermion with quadratic (instead of linear) dispersion, due to absence of Lorentz symmetry.

We now turn to the more interesting case in which there is a {\em
finite} density of fermions in the ground state $|\psi_0\rangle$. In
order to sustain this we must have a higher chemical potential for
the fermions, thus $\Delta \mu = \mu_F-\mu_B > 0$.
$\Delta\mu$ can be viewed as a chemical potential for Goldstino; in its presence the fermionic Goldstino modes are ``filled
up" to some Fermi wavevector $k_F$.

In the presence of $\Delta \mu  > 0$, $H_G$ is no longer
supersymmetric; it instead has a nonzero
commutation relation with the supersymmetry generators:
\begin{eqnarray}
&&[Q,H_G]= -[Q,\mu_F N_F+\mu_B N_B]=-\Delta\mu Q; \nonumber\\
&&[Q^\dagger,H_G]=-[Q^\dagger,\mu_F N_F+\mu_B N_B]=\Delta\mu
Q^\dagger. \label{susy1}
\end{eqnarray}
From (\ref{susy1}) we can immediately conclude the following: (i)
$Q|\psi_0\rangle$ is an {\em exact} excited state with excitation
energy $\Delta\mu$; and (ii) $Q^\dag|\psi_0\rangle=0$, because if it
were not null, it would be a state with {\em negative} excitation
energy $-\Delta\mu$. We thus find even though in this case
supersymmetry is {\em explicitly} broken by $\Delta\mu$, we still
have a {\em sharp} zero momentum fermionic collective mode generated
by $Q$, which is now {\em gapped}. The situation is somewhat similar
to what happens to a ferromagnet in an external magnetic field: the
field breaks rotation symmetry and opens a spin wave gap, but the
spin wave remains a sharp collective mode. This is a ``hole-like"
Goldstino mode since it is created by $Q$ instead of $Q^\dag$, which
creates a hole in the occupied Goldstino Fermi sea. The analogous
(to Eq. (\ref{modes})) finite momentum states are $R_{\bf
q}|\psi_0\rangle$, which are expected to have {\em downward}
quadratic dispersion of the form $E_{\bf
q}-E_0\approx\Delta\mu-\alpha|{\bf q}|^2$.

Again due to (\ref{susy1}), all eigenstates of $H_G$ except for
vacuum come in pairs whose energies differ by $\Delta\mu$ and
fermion numbers differ by one. This is because the canonical
Hamiltonian remains supersymmetric.

We now turn the discussion to possible experimental detection of the
Goldstino modes. Normally one would expect that these modes can only
be detected in processes in which a boson is turned to a fermion or
vice versa, as that is what $Q^\dag$ or $Q$ does. There is no such
process that can be easily engineered in cold atom systems that we
are aware of, except for possible co-tunneling processes in which a
fermion leaves the system and a boson enters the system
simultaneosuly, or vice versa. In the following we show that in the
presence of a Bose condensate, the Goldstino mode contributes a {\em
finite} spectral weight to the spectral function of single fermion
Green's function; as a result it can be detected through processes
that involve a {\em single} fermion. Physically this is possible
because in the presence of a Bose condensate, the boson number is
not fixed in the ground state; as a result a single fermion (hole)
can grab a boson from the condensate and propagate as the Goldstino
mode. To demonstrate this we calculate the overlap between the
fermionic single hole state $f_{{\bf q}=0}|\psi_0\rangle$ with the
normalized Goldstino mode $(1/\sqrt{N})Q|\psi_0\rangle$:
\begin{eqnarray}
&&{1\over \sqrt{N}}\langle\psi_0|Q^\dag f_{{\bf q}=0}|\psi_0\rangle
={1\over \sqrt{N}}\sum_{\bf k}\langle\psi_0|f^\dag_{\bf k}a_{\bf k}
f_{{\bf q}=0}|\psi_0\rangle\nonumber\\
&&={1\over \sqrt{N}}\left[\langle\psi_0|f^\dag_{0}
f_{0}a_{0}|\psi_0\rangle+\sum_{{\bf k}\ne
0}\langle\psi_0|f^\dag_{\bf k}a_{\bf k}
f_{{\bf q}=0}|\psi_0\rangle\right]\nonumber\\
&&=\sqrt{{N^0_B\over N}}n^f_{{\bf q}=0}+\sqrt{{1\over N}}\sum_{{\bf
k}\ne 0}\langle\psi_0|f^\dag_{\bf k}a_{\bf k} f_{{\bf
q}=0}|\psi_0\rangle, \label{spectral}
\end{eqnarray}
where $N^0_B$ is the number of bosons in the condensate and
$n^f_{{\bf q}=0}=\langle\psi_0|f^\dag_{{\bf q}=0} f_{{\bf
q}=0}|\psi_0\rangle \lesssim 1$. When $N^0_B$ is macroscopic (as is
the case in the presence of a condensate),
$a_{0}|\psi_0\rangle=\sqrt{N^0_B}|\psi_0\rangle$, and the first term
in (\ref{spectral}) dominates the second. We thus find the zero
momentum fermion Green's function has {\em finite} weight on the
Goldstino, and the weight is approximately $N_B^0/N$, proportional
to the condensate density. As a result the fermion spectral function
$A({\bf q}=0, \omega)$ has a {\em sharp} $\delta$-function peak at
$\omega=\Delta\mu$. This is highly unusual as in electronic or other
fermionic systems, one normally expects sharp (coherent) spectral
peak for fermions with momentum near $k_F$ and corresponding energy
near zero in a Fermi liquid, corresponding to Landau quasiparticles
which are well defined only near the Fermi surface. The sharp,
coherent fermion peak at {\em zero} momentum and {\em finite} energy
we find here is a remarkable consequence of the combination of
supersymmetry and Bose condensation, which is unique to such
supersymmetric Bose-Fermi mixtures. Another remarkable property is
that this spectral peak remains sharp at finite temperature $T$ as
long as $T$ is below the Bose condensation temperature $T_c$. This
is because (\ref{susy1}) guarantees that the Goldstino mode is sharp
at any $T$, while for $T < T_c$ the condensate density is nonzero,
so the fermion spectral function has a weight (proportional to
condensate density) on the Goldstino mode. We are not aware of any
fermionic mode (like Landau quasiparticles) that remain sharp at
finite $T$ in other contexts. The fermion spectral function at small
but nonzero momentum ${\bf q}$ will also have finite weight on the
finite ${\bf q}$ Goldstino mode for $T < T_c$. This will result in a
spectral peak at $\omega=\Delta\mu-\alpha|{\bf q}|^2$, whose width
goes to zero as $|{\bf q}|\rightarrow 0$.

In electronic condensed matter systems, single electron spectral
function can be measured using electron tunneling
or photo-emission. In cold atom systems we do not have equivalent
methods (yet). On the other hand alternative ways to measure single
particle Green's function or spectral function are currently being
developed, like stimulated Raman spectroscopy\cite{dao}. Once these
methods become available, the Goldstino mode with the specific
properties discussed above can be probed through the fermion
spectral function, as long as there is a Bose condensate (which is
what bosons tend to form, except at certain commensurate lattice
filling). We are thus highly hopeful that supersymmetric Bose-Fermi
mixtures can be studied and the Goldstino mode can be detected
experimentally, and our predictions can be tested.

We close by stating that we have studied possible realization of
supersymmetry in cold atom Bose-Fermi mixtures, in a {\em
non-relativistic} setting. Our study gives examples of
supersymmetry breaking, both spontaneous and explicit. In both
cases the system supports a sharp, collective fermionic excitation
which is a Goldstino-like mode; we have discussed ways to detect it
experimentally as an unambiguous signature of supersymmetry. We
hope such a study can provide hints to the mechanism of
supersymmetry breaking in relativistic quantum field theories, in
which the supersymmetry algebra is usually a graded Poincare
algebra because of the linear dispersion of the relativistic
fermion and that the boson is of the Klein-Gordon type. As a
result the anti-commutation relation between the super-charge and
its hermitian conjugate is the Hamiltonian in that case, instead
of the total particle number in the present work.

We thank Lee Chang, Xiaosong Chen, Xi
Dai, Miao Li, Hualing Shi, Zhan Xu, Zhengyu Wen, Li You, Lu Yu,
Guangming Zhang, and Zhongyuan Zhu for useful discussions, and Kavli Institute for Theoretical Physics (supported in part by the National Science Foundation under Grant No.
PHY05-51164), and Kavli Institute for Theoretical Physics China for hospitality. This
work was supported by Chinese National Natural Science
Foundation, a fund from CAS, and MOST
grant 2006CB921300 (YY), and by National Science Foundation grants No.
DMR-0225698 and No. DMR-0704133 (KY).


\begin{references}
\bibitem{note} Normally for a 1st quantized Hamiltonian, the Hilbert space is defined with {\em fixed} particle numbers. When we say that the ground state of $H$ or $H_G$ has no (or one) fermion, what we mean is among all the ground states with {\em different} numbers of bosons and fermions, this ground state has the lowest energy, which is what we just proved.
\end{references}
\end{document}